\documentclass[fleqn,10pt]{wlscirep}
\title{Density filtered Fluorescence Correlation Spectroscopy for highly concentrated solutions.}

\author[1]{Mathias Lechelon}
\author[2]{Marco Pettini}
\affil[1]{Aix-Marseille University, Centre d'Immunologie Marseille Luminy, CNRS, INSERM, 13288 Marseille, France}
\affil[2]{Aix-Marseille University, Centre de Physique Th\' eorique, CNRS UMR7332, 13288 Marseille, France}

\affil[*]{mathias.lechelon@gmail.com}


\begin{abstract}
Fluorescence Correlation Spectroscopy (FCS) is widely used to detect and quantify diffusion processes at the molecular level. The molecules of which diffusion is studied are marked with fluorescent dyes. It is commonly maintained that this technique only applies to systems where the concentration of fluorescent molecules is low. Even if this is the optimal operational condition, we show that FCS can be used also at high concentrations (up to 50$\mu$M) of fluorescent molecules: the detector blinding due to highly fluorescent solutions of concentrated dyes can be avoided by using neutral optic density (OD) filters, and the initial condition of very bad signal to noise ratio (SNR) can be hampered by suitable statistical averaging, as usual in other contexts of signal analysis. 
\end{abstract}
\begin{document}

\flushbottom
\maketitle

\thispagestyle{empty}

\section*{Introduction}
Fluorescence Correlation Spectroscopy (FCS) has been developed in the '70s \cite{webb} and rapidly became a useful technique in various fields from biology to chemistry. But FCS users get in troubles when dealing with highly concentrated solutions.
First of all, the autocorrelation functions (ACFs) tend to be squeezed as the number of fluorescent molecules in solution is increased, this fact leads to the common belief that these curves cannot be fitted any longer.
Then, a second problem is introduced by the detector, because at increasing dye-concentration it can quickly attain saturation. 
To fix these problems, some authors resorted to techniques conceived to reduce the observable volume with plasmonic nanoantennas \cite{khatua} or plasmonic gold bowtie nanoantennas \cite{kinkhabwala}.
Laurence et al. \cite{laurence} also show that the mentioned difficulties can be overcome by using several connected detectors, each one receiving part of the fluorescent beam - coming from the sample - after having separated it through beamsplitters. Thus, this setup needs many detectors and sometimes cannot be the optimal choice. An alternative, that we are putting forward here, is based on the use of absorptive filters to attenuate fluorescent light,  
and long time averaging in order to overcome the bad Signal to Noise Ratio (SNR).
 
 The present work has been motivated by a practical problem of potentially great impact: the experimental confirmation or refutation of the possibility of activating long-range electrodynamic attractive forces between biomolecules. These forces could play a relevant role in the recruitment at a distance of the partners of biochemical reactions  \cite{pre3} in living matter, besides Brownian diffusion and standard short-range forces (covalent bonds, van der Waals, and so on).

In preliminary studies - of theoretical and numerical kind  \cite{pre1,pre2} , respectively -  Fluorescence Correlation Spectroscopy is the experimental technique identified to investigate whether the mentioned electrodynamic forces could be at work in suitable conditions. This technique has to be applied to the investigation of the diffusion behavior of biomolecules in solution at different values of their concentration (that is, when the average intermolecular distance is varied). In a recent paper \cite{pre4}, a successful experimental assessment of this method was carried out by working with molecules interacting through electrostatic forces, and at a standard low level of fluorescence,   leading to the conclusion that the FCS technique is a reliable experimental procedure for an assessment of the strength of long-range intermolecular interactions. This suggests that the method can also be applied for the detection of the electrodynamic intermolecular interactions mentioned above. However, a fourth crucial step of this feasibility study remains to be investigated, that is, if FCS can still be used at high levels of fluorescence, because these will be the typical operating conditions in the experiments aimed at detecting electrodynamic intermolecular forces. Hence, the topic on which the present experimental work focuses. The outcomes, as discussed throughout the paper, clearly show that FCS can be used also at high concentrations of fluorescent molecules, that is in a-priori very bad conditions of signal to noise ratio.

\section*{Results}
Working at high density of fluorescent molecules entails an over saturation of the detector and a strong reduction of the ratio between the variance of  fluorescence fluctuations and their average. The use of Optical Density (OD) filters along the fluorescence optical path then appears as a possible way to fix the problem of detector blinding. The drawback being that density filters
randomly absorb photons, and, since within the very short sampling time lags the number of photons is relatively small, this is a source of noise deteriorating the autocorrelation property of the fluorescence signal.  However, as we shall see in the following, by resorting to the standard method of increasing the statistics of the signal acquisition, the SNR can be conveniently improved allowing to retrieve the desired information. 
\subsection*{FCS measurements with filters}
With a first experiment,  by means of FCS, we measured the diffusion coefficient of the fluorescent dye Atto 488 (AT488) solvated in water at 1 nM concentration. This dye has a strong absorption peak at 500 nm and high fluorescence quantum yield peaked at 520 nm. The measurements have been carried out both with and without the OD filters. The transmission coefficients of the OD filters that we adopted were: 10$\%$ (OD1), 5$\%$  (OD1.3) and 1$\%$  (OD2), respectively.
We have also recorded the background noise obtained without the solution and with the laser switched off.

\begin{figure}[h!]
\centering
\includegraphics[width=0.8\linewidth]{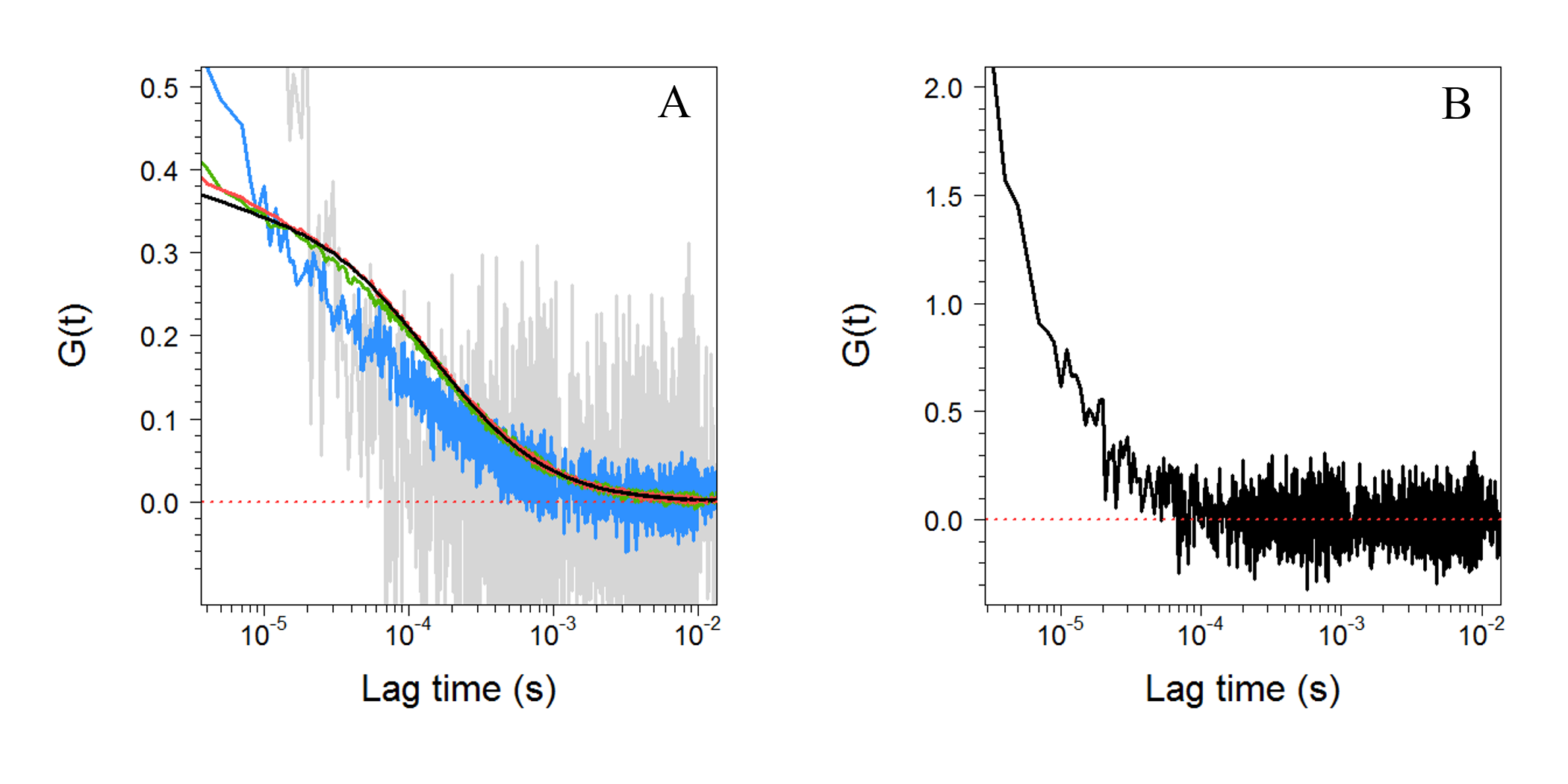}
\caption{FCS results. Panel A displays the ACFs obtained for a 1nM solution of AT488 and: without filter (black curve), with OD1 filter (red curve), with OD1.3 filter (green curve), with  with OD2 (blue curve). The ambient noise has also been recorded and its ACF has been drawn in light grey. Panel B displays the ACF of the ambient noise. }
\label{fig:Figure1_FCS+filtres+problems_2}
\end{figure}

Figure \ref{fig:Figure1_FCS+filtres+problems_2} summarizes the outcomes of the above mentioned measurements. Figure \ref{fig:Figure1_FCS+filtres+problems_2}A displays the autocorrelation curves obtained with the AT488 solution, without filter (black curve), with the filters described above (red, green and blue for OD1, OD1.3 and OD2 respectively) and the background noise (light grey). Higher transmission coefficient filters (OD1 and OD1.3) show an excellent agreement with measurements performed without OD filters, and the ACFs overlap. 
To the contrary, the lower transmission coefficient filter (OD2) yields an ACF which is more similar to the background noise curve. This discrepancy is evidently due to the strong attenuation of the fluorescence operated by the OD2 filter which makes the SNR very poor by letting only $1\%$ of the signal arrive to the detector. 

Also the background noise has been autocorrelated and the outcome is displayed in Figure \ref{fig:Figure1_FCS+filtres+problems_2}B. In the absence of any bias, the autocorrelation of the background noise should look like a delta function with a flat noisy residue of zero average. But  this is not the case of the ACF in Figure \ref{fig:Figure1_FCS+filtres+problems_2}B which is rather typical of the so called and well-known afterpulsing phenomenon \cite{zhao2003}. The afterpulsing is a non-ideal individual behavior of single-photon avalanche diode detectors - like the one used in this study - affecting the measures by adding to each real signal pulse an afterpulse at a later time  \cite{Ziarkash2018}. 

\subsection*{FCS simulations}
\begin{figure}[h!]
\centering
\includegraphics[width=0.7\linewidth]{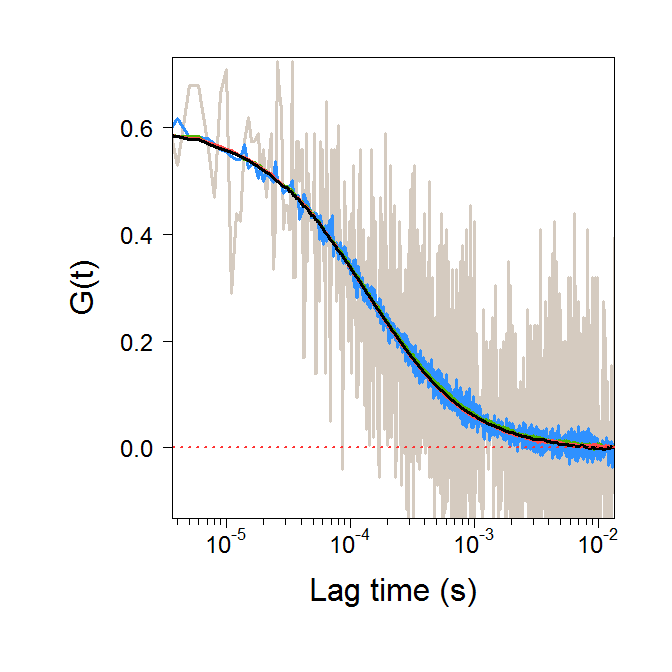}
\caption{Numerical FCS simulations. ACFs from numerical simulations modelling freely diffusing particles. The parameters have been chosen so as to reproduce the experimental  conditions. The ACFs displayed have been obtained from a simulated solution of 1nM of fluorescent particles, with: no filter (black),  OD1 filter (red), OD1.3 filter, OD2 filter (blue) and OD3 (light brown).}
\label{fig:Figure2_FCS-simulations}
\end{figure}

In order to independently check and better understand our first results, we have performed numerical FCS simulations considering parameters that reasonably reproduced the experimental conditions. We have numerically simulated the diffusion of 1nM of fluorescent particles in solution, and the signal acquisition has been made by mimicking the presence of OD filters. More details are given in Methods section. We have used the following geometry and simulation parameters: a cubic container with 10$\mu$m long edges, filled with a solution of 1nM of fluorescent particles, what corresponds to 602 molecules. Based on the experimental measurements we have set the diffusion coefficient of the particles to 408$\mu m^2/s$ to comply with AT488 diffusion coefficient at 20$^{ o}$C, and the number of photons emitted per particle and per acquisition time has been fixed at $n_{ph}$=5. The OD filters used in the experiments have also been simulated, with 10$\%$, 5$\%$, 1$\%$ transmission (corresponding to OD1, OD1.3 and OD2 filters respectively), also the case of a very opaque filter has been simulated with a transmission of 0.1$\%$ (OD3). \\
The results are reported in Figure \ref{fig:Figure2_FCS-simulations}. Unlike with the experimental results shown in Figure \ref{fig:Figure1_FCS+filtres+problems_2}A, also the ACF obtained with the OD2 filter overlaps with the other ACFs. As expected, the higher the OD filter, the noisier the ACF. Notice that the light grey ACF in Figure \ref{fig:Figure1_FCS+filtres+problems_2}A representing the ambient noise autocorrelation is missing in Figure \ref{fig:Figure2_FCS-simulations} because to simulate the ambient noise we had to simulate and correlate a random signal, and this entails an a-priori known ACF, that is, for a white noise a delta-like autocorrelation with a flat, noisy, zero-average pattern at any time delay. \\
These results thus show that simple simulations of fluorescent diffusing particles detected in a confocal volume with different  OD filters give very similar ACF shapes leading to equal diffusion times after fitting. This proves that the OD filters add noise to the ACFs, but do not affect the ACFs shape and consequently the diffusion time of the particles. Moreover this confirms that the phenomenon observed in Figure \ref{fig:Figure1_FCS+filtres+problems_2}B is an experimental artefact due to the afterpulsing. 

\subsection*{FCCS measurements with filters}
To get rid of the ACFs deterioration due to the previously mentioned afterpulsing of the detector, we have run other experiments aimed at circumventing this artefact.  The main techniques used for this purpose are Fluorescent Lifetime Correlation Spectroscopy (FLCS) and Fluorescence Cross-Correlation Spectroscopy (FCCS). Enderlein et al. \cite{Enderlein2005} have shown that FLCS can correct the correlation curves affected by this artefact, by using time-correlated single-photon counting in order to separate the afterpulsing events from the true fluorescent signal. The other technique, called Fluorescent Cross-Correlation Spectroscopy (FCCS), can avoid afterpulsing \cite{Widengren1995} by splitting the fluorescent signal with the use of a 50/50 beamsplitter and sending the two signals to two independents detectors, then by cross-correlating these signals. We have chosen to adopt the FCCS technique, and two experiments have been performed: the first one at a constant concentration with different OD filters; the second one with different concentrations (from 1nM to 50$\mu$M) with suitable OD filters.

\subsubsection*{At a constant concentration}
{Several experiments have been performed at the constant concentration of 100 nM aqueous solution of AT488 molecules,  and using different OD filters in order to figure out their effects on the correlations curves. This concentration has been selected to fulfil both of the following conditions: the detectors are not blinded in the absence of OD filters, and the SNR is high enough for experiments performed with the highest OD filter used (OD2). The results of experiments performed with the FCCS device are displayed in Figure \ref{fig:Figure3_FCCS+filtres+ACFs} where, in panel A, those worked out in the absence of OD filters are represented by the black curve, those obtained with the OD1 filter are represented by the red curve, those  with the  OD1.3  filter by the green curve,  and those with the OD2 filter with the blue curve. All the Cross-Correlation Functions (CCFs) overlap very well even if with increasing fluctuations around the reference curve (the black one) at increasing optical density of the filter. These outcomes show some discrepancy with respect to those in Figure \ref{fig:Figure1_FCS+filtres+problems_2}, in particular at short time lags and in the case of the OD2 filter, whereas there is a closer agreement with the outcomes of the numerical simulations reported in Figure \ref{fig:Figure2_FCS-simulations}. To confirm that the afterpulsing has been eliminated through the cross-correlation analysis, individual ACFs obtained from each individual detector are displayed on panels B and C (detector 1 and 2 respectively). 
\begin{figure}[h!]
\centering%
\includegraphics[width=0.9\linewidth]{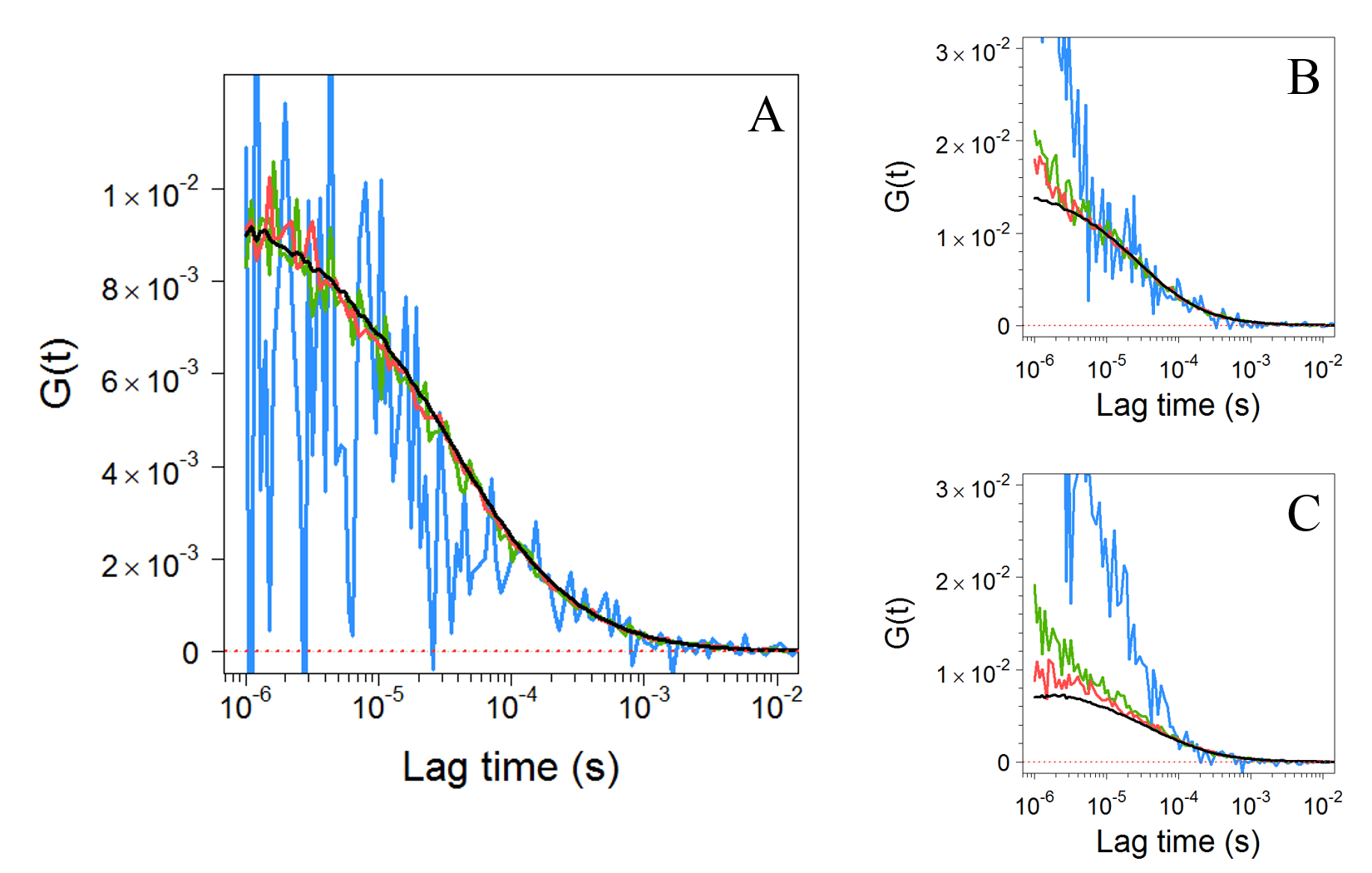}
\caption{FCCS results. Panel A displays CCFs obtained without filter (black),with OD1 filter (red), OD1.3 filter (green) and OD2 filter (blue). Panel B and C display the related ACFs with the same color conventions for detectors 1 and 2 respectively.}
\label{fig:Figure3_FCCS+filtres+ACFs}
\end{figure}
Clearly, these ACFs are  strongly affected by the afterpulsing occurring in each detector. The relevance of the afterpulsing increases with the OD value. The overlapping CCFs shown in Figure \ref{fig:Figure3_FCCS+filtres+ACFs}A are also individually displayed in Figure \ref{fig:Figure3-2_FCCS+filters-separated} with the corresponding fittings and residuals. For a better comparison, the residuals are displayed in percentage with 100$\%$ corresponding to $G_0$. As expected, the higher the OD filter, the noisier the residuals.\\
\begin{figure}[h!]
\centering%
\includegraphics[width=0.97\linewidth]{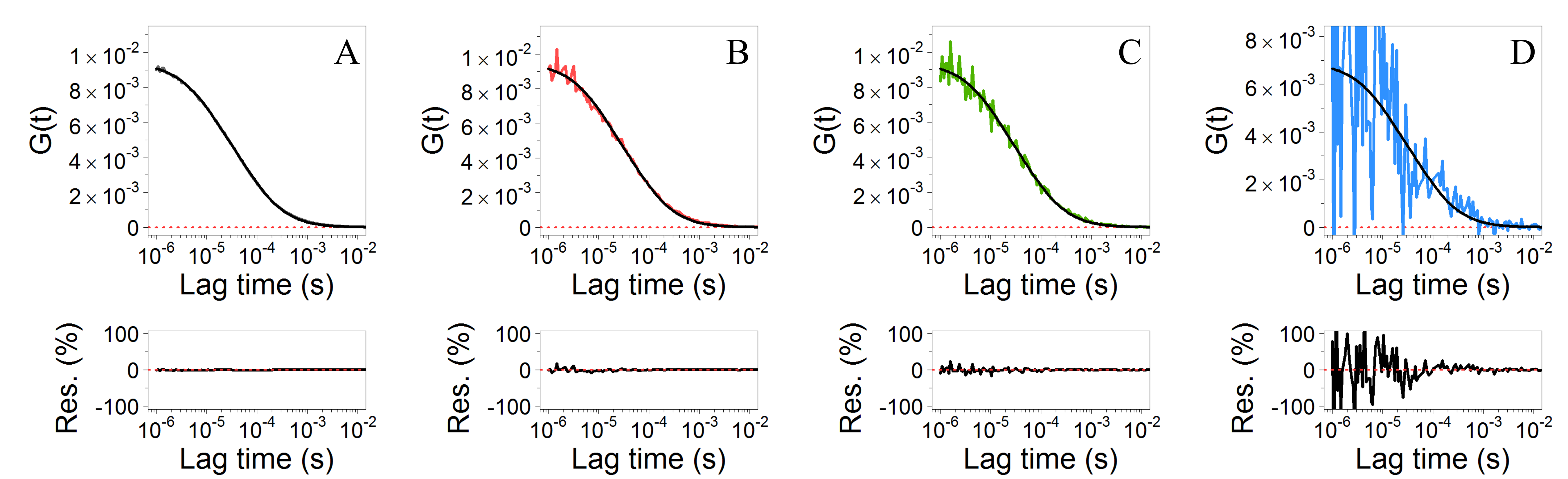}
\caption{FCCS results. Panels from A to D display the CCFs and their fitting from measurements performed on 100 nM solutions without OD, with OD1, with OD1.3 and with OD2. Both the triplet time and the fraction of molecules in the triplet states have been kept equal to the values reported on fitting from panel A (with no filter). Below each CCF, the residuals are displayed in percentage, 100$\%$ corresponding to $G_0$.}
\label{fig:Figure3-2_FCCS+filters-separated}
\end{figure}

\begin{figure}[h!]
\centering%
\includegraphics[width=0.9\linewidth]{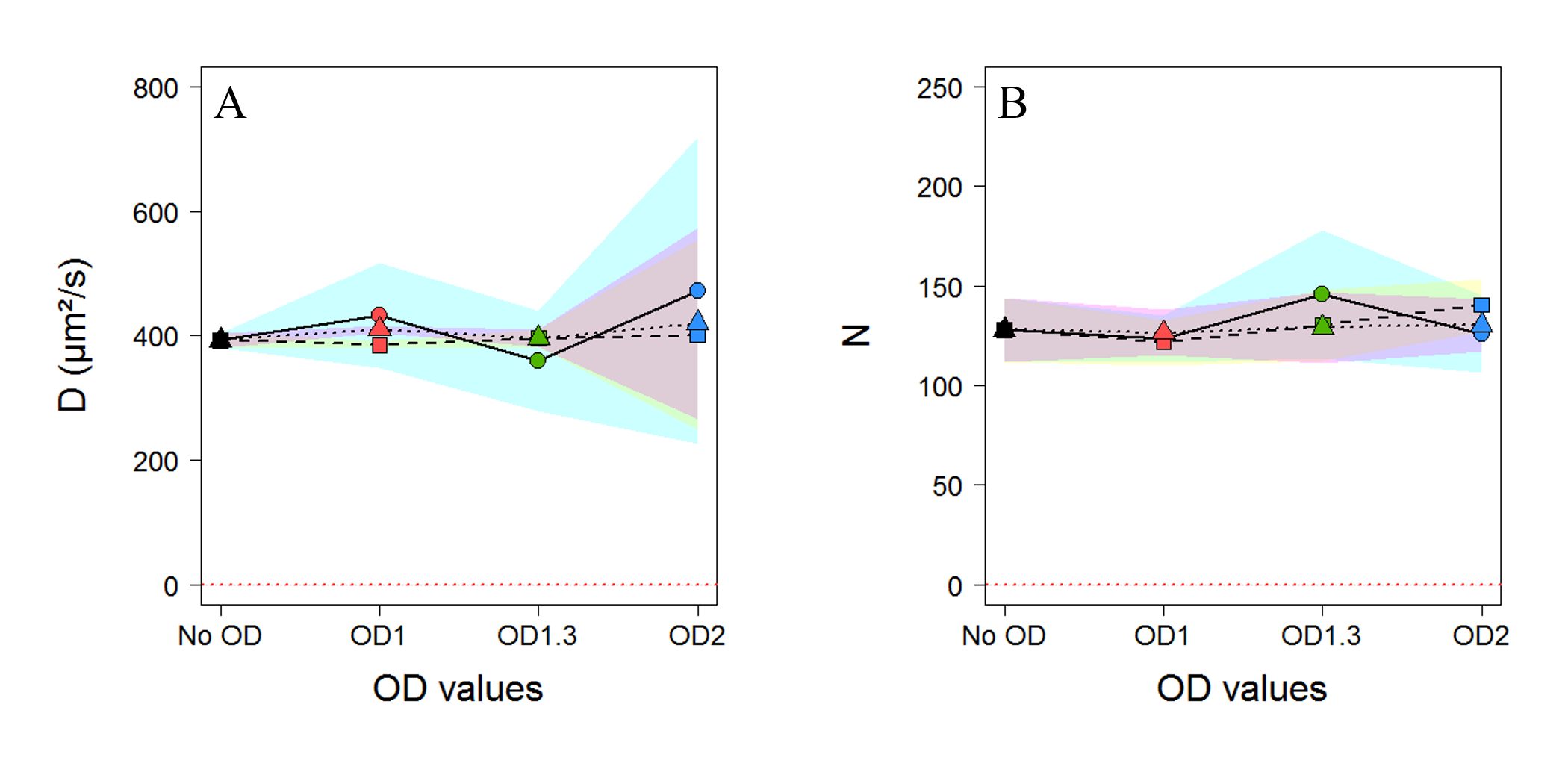}
\caption{FCCS results. Panel A displays the diffusion coefficients and panel B the estimated number of molecules in the confocal volume calculated for a solution of 100nM; measurements have been performed without OD filter (black symbols), with OD1 filter (red symbols), with OD1.3 filters (green symbols) and with OD2 filter (blue symbols). The CCFs obtained have been fitted with different parameters: with free triplet time and free triplet state fraction (disks and cyan error bars), with fixed triplet time and free triplet state fraction (squares and magenta error bars), with fixed triplet time and fixed triplet state fraction (triangles and yellow error bars). }
\label{fig:Figure5new_N+D_2}
\end{figure}

The fittings shown in Figure \ref{fig:Figure3-2_FCCS+filters-separated} have been performed by taking the triplet time $\tau_T = 9.7\mu s$, and the fraction of molecules in the triplet state
Equal to $22\%$, obtained from the CCF of panel A and then kept fixed in the fittings of the other CCFs. The acquisition time for each of the CCFs reported in panels A, B, and C was of almost one hour divided in 72 intervals of 50 seconds. Then each final CCF - in panels A, B, and C - is obtained by averaging on the set of 72 CCFs corresponding to each interval.
The overall acquisition time, as well as the number of intervals, for the CCF reported in panel D was doubled. Three different fitting settings have been tested. In a first place the following parameters have been kept free: diffusion time, number of molecules, time spent in the triplet state and the fraction of the molecules in the triplet state (circles in the figure). As is observed in Figure \ref{fig:Figure5new_N+D_2}A, while the average diffusion coefficients are in close agreement for all the measurements performed with OD filters, but the error bars are sensitive to the OD filter value (cyan). CCF fitting has also been performed with the fraction of the molecules in the triplet state free and the time spent in the triplet state fixed (squares), and with both of these parameters fixed (triangles). These two fitting settings give similar results as is seen in Figure \ref{fig:Figure5new_N+D_2}A, for what concerns both the average diffusion coefficient and the error bars (magenta and yellow respectively). A synopsis of the average outcomes with the corresponding error bars  is given in Figure \ref{fig:Figure5new_N+D_2}. The estimated number of molecules inside the confocal volume according to the different fitting settings are also visible in Figure \ref{fig:Figure5new_N+D_2}B. In this figure, the number of molecules remains stable for measurements performed with any filter, as expected, due to the constant concentration of the fluorescent solution. The error bars are also reaching similar values for the three fittings performed, which differ from the diffusion coefficient observed on panel A.

Then, working again at the constant concentration of 100 nM of AT488, and performing all the measurements with the OD2 filter, the effect of the averaging time has been checked.  In Figure \ref{fig:Figures6_IncreaseAverage}, four examples of CCFs obtained at different averaging times are reported. In Figure \ref{fig:Figures6_IncreaseAverage}E the diffusion coefficient is worked out after fitting at different times elapsed  during the experiment, ranging from 12 minutes to 144 minutes (every 12 minutes), and adopting the three fitting settings above described. 

As observed in this figure, the results obtained from these three fitting settings have a similar trend but with noticeable differences. The fitting performed with fixed triplet time and fixed triplet fraction (triangles) gives a diffusion coefficient which - after sufficient averaging - is the closest to the one measured without OD filter. These results are also reported in Figures \ref{fig:Figure5new_N+D_2}A and \ref{fig:figure8-2_NetD}A, where it can be seen that the corresponding error bars are small compared to the other diffusion values worked out with different fitting settings. The results obtained with fixed triplet time and free triplet fraction (squares) after fitting lead to diffusion coefficients higher than those obtained with fixed triplet state and fraction (with an increased value of about 50$\mu m^2/s$ reaching even 200$\mu m^2/s$), and follow in parallel the same pattern of the triangles. This is due to the triplet fractions estimated at 0 or close to 0 (data not shown) after fitting, as a consequence the estimates of the diffusion time of the AT488 are consequently modified and lowered because of the missing triplet contribution. Finally, the fitting performed with both free triplet state and fraction gives diffusion coefficients following the same trend than the other fittings but with oscillating values. The triplet fractions and triplet times estimates have been found covering wide ranges from 0 to 1 for the triplet fraction and 0$\mu s$ to 97$\mu s$ for the triplet time (data not shown), which modifies the estimated diffusion time of AT488. The differences observed between these fittings are essentially due to the rather high noise observed on the correlation functions, and by fixing the triplet state parameters (fraction and time), the fitting results are improved.
\begin{figure}[h!]
\centering%
\includegraphics[width=0.9\linewidth]{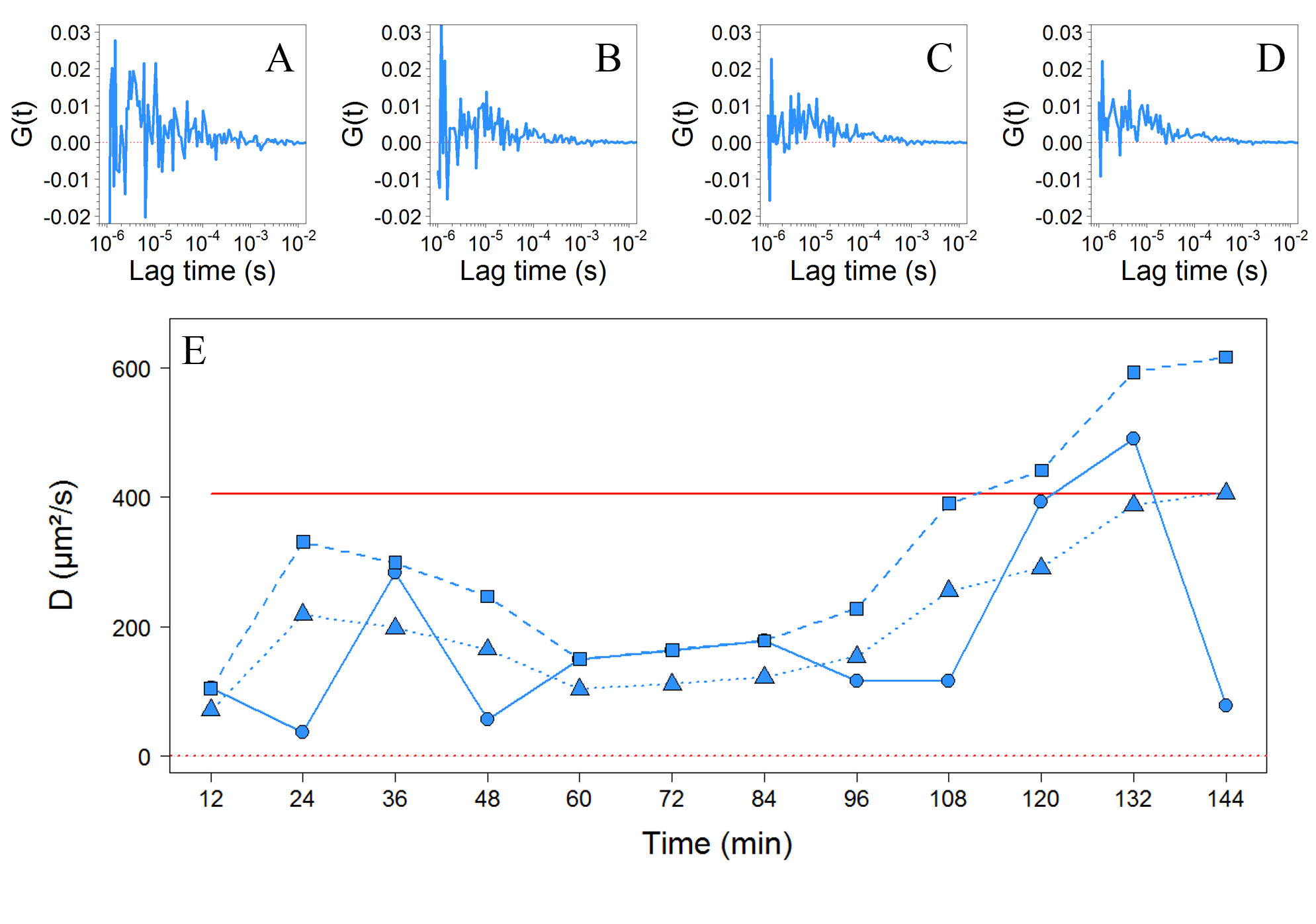}
\caption{FCCS results. Panels from A to D display CCFs obtained from measurements on a 100 nM solution, after 12 minutes, 48 minutes, 96 minutes and 144 minutes respectively. The diffusion coefficients D obtained from the fitting on the CCFs at different time measurements are visible on panel E. D from fitting with free triplet time and triplet state fraction are displayed with circles, with fixed triplet time and free triplet state fraction with squares and with fixed triplet time and fixed triplet state fraction with triangles. }
\label{fig:Figures6_IncreaseAverage}
\end{figure}


\subsubsection*{At varying concentrations}
In order to complement our investigation of the effects of OD filters in fluorescence correlation measurements, experiments have been also performed with different  concentrations (1nM, 1$\mu$M, 10$\mu$M and 50$\mu$M) and different filters chosen so as to avoid detector blinding (no OD filter, OD1, OD1.3 and OD2, respectively). Figure \ref{fig:Figure7_FCCS+filters} displays the results of the experiments. In panel A, the CCFs obtained with high concentrations (1$\mu$M, 10$\mu$M and 50$\mu$M in red, green and blue respectively) are squeezed at such a point that they are not visible compared to the CCF obtained at low concentration (1nM in black), because the maxima of the CCFs are inversely proportional to the number of molecules $N$. This often leads FCS users to the common assumption that ACFs or CCF from highly concentrated solutions cannot be fitted properly. Therefore, panel B displays the same CCFs reported in panel A but rescaling their values just by multiplication with the number $N$ of molecules. From panel B, one can observe the overlapping of the different CCFs so rescaled, what provides a first overview of the good agreement among the shapes of the CCFs obtained for the samples at different concentrations with the respective filters. \\
\begin{figure}[h!]
\centering%
\includegraphics[width=0.9\linewidth]{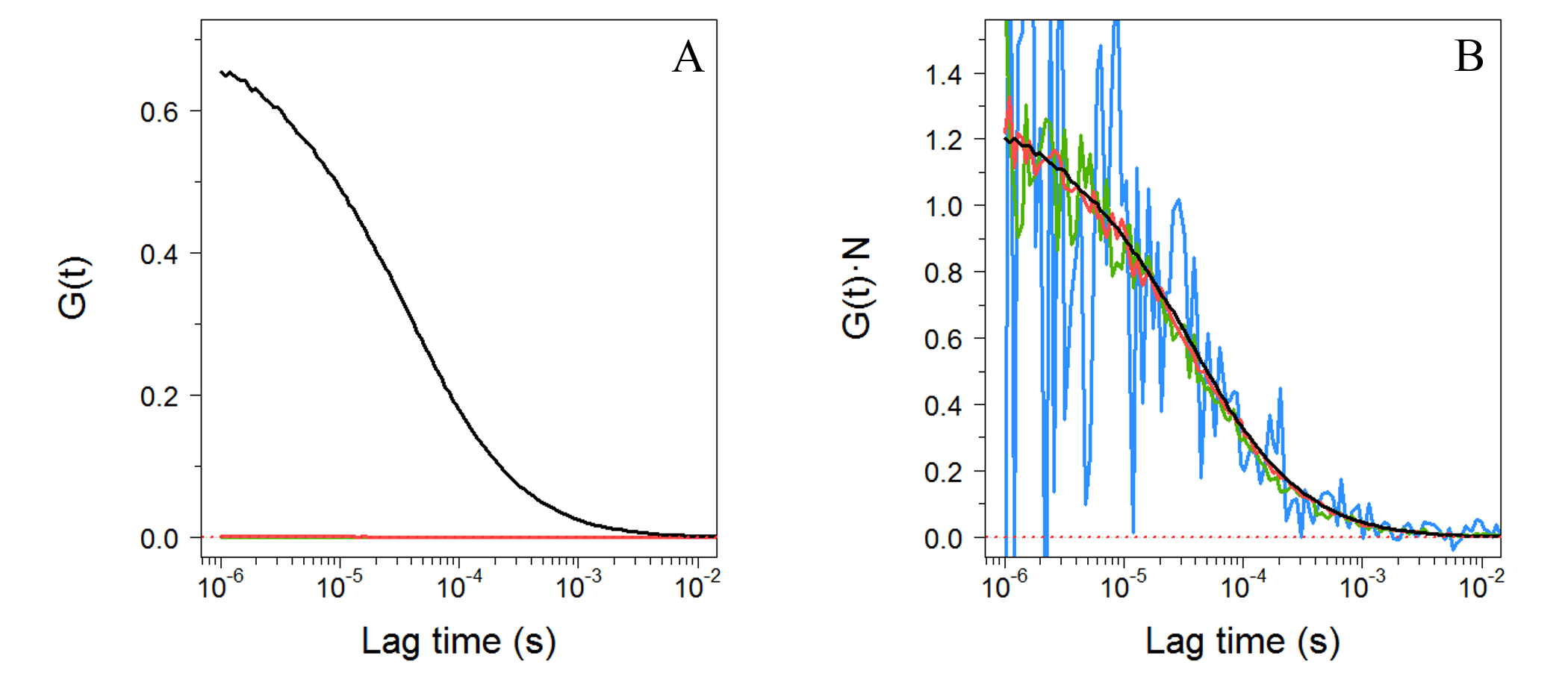}
\caption{FCCS results.On panel A, CCFs are visible obtained from solutions of 1nM, 1$\mu$M, 10$\mu$M and 50$\mu$M (in black, red, green and blue respectively). On panel B these same CCFs have been normalized with the number of molecules N estimated in the confocal volume.}
\label{fig:Figure7_FCCS+filters}
\end{figure}
Figure \ref{fig:Figure7-2_FCCS+filtres-separated} displays the same results of Figure \ref{fig:Figure7_FCCS+filters}, with the individual CCFs, their fitting, and the corresponding residuals in percentage. As already observed in Figure \ref{fig:Figure3-2_FCCS+filters-separated}, the higher the OD value, the noisier the CCF. \\
\begin{figure}[h!]
\centering%
\includegraphics[width=0.9\linewidth]{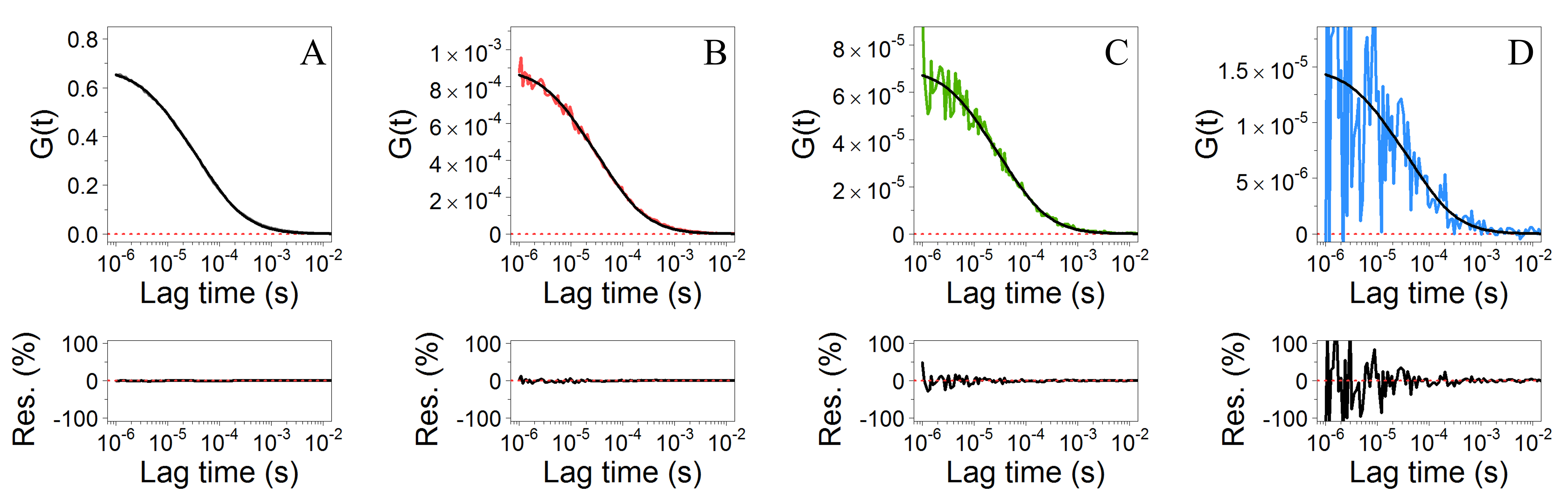}
\caption{FCCS results. Panels A to D display the CCFs visible in Figure \ref{fig:Figure7_FCCS+filters}, with solutions of 1nM, 1$\mu$M, 10$\mu$M and 50$\mu$M respectively. The fitting curves are also displayed in black on each panel, and the residuals are visible below each CCF (displayed in percentage as previously mentioned on figure \ref{fig:Figure3-2_FCCS+filters-separated}).}
\label{fig:Figure7-2_FCCS+filtres-separated}
\end{figure}
Finally, the results of these fittings performed  at varying concentrations, and with the OD filters, have been reported in Figure \ref{fig:figure8-2_NetD}. As observed in panel A of Figure \ref{fig:figure8-2_NetD}, the diffusion coefficients obtained after each of the three fittings adopted show stable results obtained at any concentration. The error bars also increase with the concentration and the OD filter value. In comparison with Figure \ref{fig:Figure5new_N+D_2}A, the error bars are definitely smaller,  reduced by several tens of $\mu$m$^2$/s for measurements performed with high values of the OD filters. These results are attributed to the high concentrations of the samples used in the experiments (50$\mu$M with OD2 filter) yielding a higher photon rate recorded by the detectors, compared to the previous experiment performed at the concentration of 100nM, bringing about much lower fluorescence. Concerning the number of molecules estimates, which are displayed on panel B, one can see that $N$ linearly increases with the concentration. After fitting, a power regression $y=ax^{b}$ is found, with $a=1.7$ and $b=1$. As explained by R\"{u}ttinger et al. \cite{RTTINGER2008}, this results also serve to determine the effective volume of the FCS device, using samples with known concentrations. The linear dependence between the average number of particles and the concentration observed in the current experiments allow to calculate the effective volume out of the resulting slope.

\begin{figure}[h!]
\centering%
\includegraphics[width=0.9\linewidth]{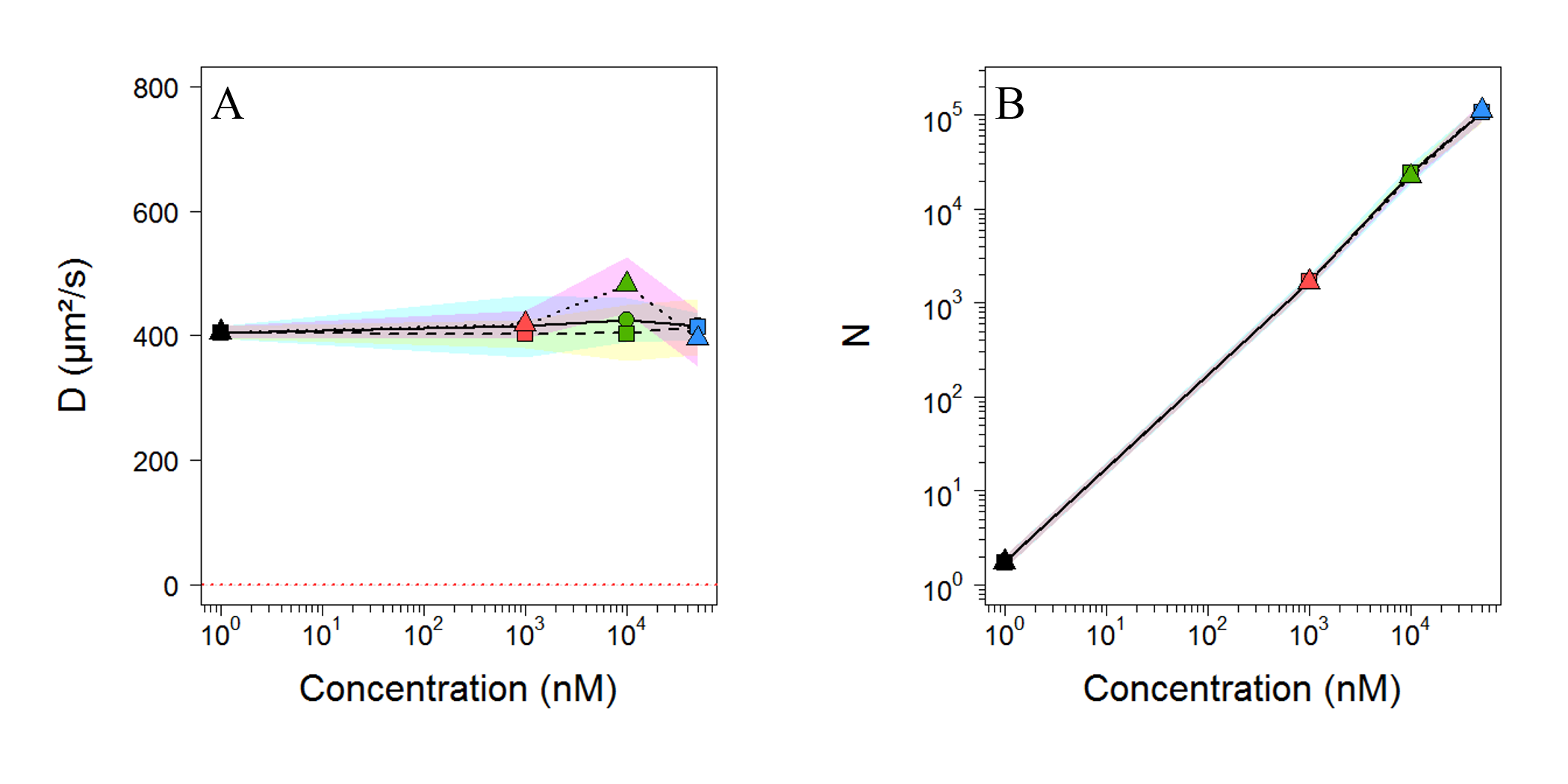}
\caption{FCCS results. Panel A displays the diffusion coefficients and panel B the estimated number of molecules in the confocal volume calculated for solutions of 1nM, 1$\mu$M, 10$\mu$M and 50$\mu$M; measurements have been performed without OD filter (black symbols), with OD1 filter (red symbols), with OD1.3 filters (green symbols) and with OD2 filter (blue symbols). The CCFs obtained have been fitted with different parameters: with free triplet time and free triplet state fraction (disks and cyan error bars), with fixed triplet time and free triplet state fraction (squares and magenta error bars), with fixed triplet time and fixed triplet state fraction (triangles and yellow error bars). }
\label{fig:figure8-2_NetD}
\end{figure}



\section*{Discussion}
In this paper we have explored FCS experiments performed at high concentrations, up to 50$\mu$M. Similar researches have been conducted by different groups \cite{khatua, laurence, kinkhabwala} confirming the interest of performing FCS experiments at high concentrations. The present work has been motivated by the need to tackle a biophysical problem requiring to measure the diffusion coefficient of fluorescently labeled  biomolecules at high concentrations \cite{pre1,pre2,pre3,pre4}. We have resorted, on the experimental side to the use of optical density filters placed along the optical path of the outgoing fluorescence beam, and on the data treatment side to a substantial increase of the statistics.

As stated by Gregor et al.\cite{Gregor2005}, the statistical accuracy of an FCS measurement roughly scales with the square of the fluorescent rate. According to this assumption and considering a 1 minute measurement of a fluorescent dye at fixed concentration, the time needed to make similar measurements with OD1, OD1.3 and OD2 filters would be approximately of 100 minutes, 2000 minutes and 10000 minutes respectively. Experimentally, we have been able to make proper measurements with correct fittings with measurements lasting two hours when using the OD2 filter, and lasting one hour when using OD1.3 and OD1 filters. \\
While the technique that we have used might not be suitable for every biological sample containing moving objects, due to the long time required to perform the measurments, it can be used to measure the diffusion coefficients of biomolecules in highly concentrated solutions.
\begin{table}[h!]
\centering
\begin{tabular}{|l|l|l|l|}
\hline
Article & Technique & Concentration \\
\hline
Khatua et al.\cite{khatua} & Single gold nanorod & 1$\mu$M\\
\hline
Kinkhabwala et al.\cite{kinkhabwala} & Gold bowtie nanoantennas & 1$\mu$M\\
\hline
Laurence et al.\cite{laurence} & APD banks & 38$\mu$M \\
\hline
This work & OD filters & 50$\mu$M \\
\hline
\end{tabular}
\caption{\label{tab:example}Comparison between FCS techniques to work with highly concentrated solutions}
\end{table}

\section*{Methods}
\subsection*{Experimental setup}

Experiments have been performed using aqueous solutions of Alexa Fluor 488 (AF488) in order to carry out the waist size calibration, and aqueous solutions of Atto 488 dye (AT488)  at different concentrations, that is, 1 nM, 1$\mu$M, 10$\mu$M and 50$\mu$M, and with different optical filters of density 2.0 (OD2, $1\%$ transmission), 1.3 (OD1.3, $5\%$ transmission) and 1.0 (OD1, $10\%$ transmission). The aqueous solutions of the AF488-dye were put in 8 wells Labtek supports to prevent evaporation.\\
{FCS measurements were done by means of a custom-made apparatus using an Axiovert 200 M microscope (Zeiss, Germany) with an excitation 488 nm Ar$^{+}$-ion laser beam focused through a Zeiss water immersion Apochromat 40X/1.2 numerical aperture objective. The fluorescence was collected by the same objective, separated from the excitation light using a dichroic mirror, and then delivered to an avalanche photodiode (SCPM AQR-13, Perkin Elmer) through 545/20 nm bandpass filter. A 50 $\mu$m diameter confocal pinhole reduced the out-of-focus fluorescence. Prior to the measurements the system has been switched on for about  60 min  in order to attain the stabilization of all the 
components. The laser waist $ \omega_{x,y}$ was set by selecting with a diaphragm the lateral extension of the laser beam falling onto the back-aperture of the microscope objective \cite{Billaudeau2013} and was then estimated using  the diffusion of Alexa FLuor 488 in water  $\omega_{x,y}=\sqrt{4 D \tau_D}$. The diffusion coefficient for AF488 available in the literature \cite{Petrek2008} ($D_{AF488,22.5^{\circ}C}= 435 \mu$m$^2/$s) has been corrected to account for the temperature at which we have operated ($20^{\circ}$C) and to account for the value of the viscosity of water at the same operational temperature (Eqs.\eqref{stokes} and \eqref{eta}), giving $D_{AF488,20^{\circ}C} = 406 \mu$m$^2/$s. We used a power of $100 \mu$W at the back-aperture objective for both AF488 and AT488 dyes.}
{FCCS measurements were performed on a commercial FCS setup (ALBA FCSTM, from ISS Inc., Champaign, America) with two excitation picosecond/CW diode lasers operating at 488 and 640 nm (BDL-488-SMN, Becker and Hickl, Germany) with a repetition rate of 80 MHz, focused through a water immersion objective (CFI Apo Lambda S 40X/1.25 WI, Nikon). The fluorescence was collected by the same objective, splitted into two detection paths by a 50/50 beam splitter (Chroma 21000) and filtered by two Emission filters (525/40 nm band pass, Semrock FF02-525/40  and 675/67 nm band pass, Semrock FF02-675/67 for the green and red channels, respectively) and detected by two avalanche photodiodes (SPCM AQR-13 and SPCM ARQ-15, Perkin Elmer / Excelitas).}
{\subsubsection*{Autocorrelation, corss-corelation and data treatment} 
\label{acf-data}
The autocorrelation function $G(\tau)$, originated by molecules  diffusing in and out of the observation volume, is defined by 
\begin{equation}
G(\tau) = \dfrac{\langle \delta F(t) \delta F(t+\tau)\rangle}{\langle F(t)\rangle^2}
\label{ACFexp}
\end{equation}
 where $ \langle F(t)\rangle $ is the average intensity, $\delta F(t)$ the intensity of fluctuations, and the brackets mean time average.\\
Similarly the cross-correlation function, obtained with the use of two independent photo-detectors, is defined by 
\begin{equation}
G(\tau) = \dfrac{\langle \delta F_1(t) \delta F_2(t+\tau)\rangle}{\langle F_1(t)\rangle \langle F_2(t)\rangle}
\label{CCFexp}
\end{equation}
with $\delta F_1(t)$ and $\delta F_2(t)$ the fluorescence intensity fluctuations obtained from detectors 1 and 2 respectively.\\
The general procedure consists in fitting $G(\tau)$ with the appropriate mathematical model describing the characteristics of the system under study. The analytical form of the autocorrelation function  for a single molecular species, assuming a three-dimensional Gaussian profile of the excitation beam accounting for diffusion \cite{Magde1974} and a triplet state of the dye \cite{Widengren1995}, is :
\begin{equation}\label{ACF}
G(\tau)=1+\dfrac{1}{N} \,\dfrac{1+n_{T}\,\exp{\left(-\dfrac{\tau}{\tau_{T}}\right) } }{\left(1+\dfrac{\tau}{\tau_{D}}\right)\sqrt{1+s^{2}\dfrac{\tau}{\tau_{D}}}} \;.
\end{equation}
Here $N$ stands for the number of molecules in the FCS observation volume, $\tau_{D}$ is the diffusion time through this volume, $ \tau_{T} $ the  triplet lifetime, $n_{T}= Tr/(1-Tr)$, with $Tr$ the fraction of molecules in the triplet state.
The dimensionless parameter $s$, called structure parameter, describes the spatial properties of the detection volume. It is given by $s=\omega_{x,y} / \omega_{z}$, where the parameter $\omega_{z}$ is related to the length of the detection volume along the optical axis, and the radial waist $\omega_{x,y}$ is related to the radius of its orthogonal section.
The diffusion coefficient $D$ is expressed as a function of the radial waist $\omega_{x,y}$, and of the diffusion time $ \tau_{D}$ by:
 \begin{equation}\label{diffusion}
 D=\omega_{x,y}^{2}/4\tau_{D} \;,
 \end{equation}
and for isolated molecules following a  Brownian motion, the hydrodynamic radius $R_H$ may be computed using the Stokes$-$Einstein equation:
\begin{equation}\label{stokes}
R_{H}=\dfrac{k_B T}{6\pi \eta(T) D} \;,
\end{equation}
where $T$  is the absolute temperature, $k_B$ the Boltzmann constant, and $\eta$ the viscosity of the fluid. The viscosity of liquids is a decreasing function of temperature and is expressed empirically between $0^{\circ}$C and $370^{\circ}$C, with an error of 2.5 $\%$, by the expression \cite{AL-SHEMMERI1993}
\begin{equation}\label{eta}
\eta(T) = A \times 10^{B/(T - C)} \;.
\end{equation}
For water, the parameters $A, B$ and $C$ are equal to $2.414\times 10^{-5}$ Pa s, 247.8 K and 140 K, respectively.}

\subsection*{Data treatment}
FCS experiments have been made with 60 measurement of 30 seconds each. Raw data have been exported as csv files with an temporal resolution of $10^{-7}$ seconds. The autocorrelation function has been obtained by means of a Fast Fourier Transform algorithm as:\\
$G(t)=ffti(fft(trace)*Conj(fft(trace)))/(sum(trace)^2-1)$\\
Where \textit{trace} stands for the fluorescence vector recorded during the acquisition time, \textit{fft} stands for the operation of fast Fourier transformation, \textit{ffti} for the inverse operation of fast Fourier transformation, \textit{Conj} means just taking the complex conjugate of the signal, and $\star$ stands for the convolution product.
All the individual curves obtained in a run have been averaged to enhance the SNR.\\
FCCS experiments have been done with 72 measurements of 50 second each (for a total of 1h) when no filters were used, when using OD1 filters and OD1.3 filters. Experiments performed with OD2 filters have been run for 144 measurements of 50 seconds.

\subsection*{Simulation of diffusion and its detection}
\subsubsection*{Scheme of numerical simulations}
Numerical simulations have been done by borrowing the code from Wawrezinieck et al. \cite{Wawrezinieck_2005} and adapting it to work in three dimensions. A virtual cube is created with an edge size of $d$=10${\mu}m^2$; periodic boundary conditions are assumed. The box contains $n$ independant particles moving randomly in order to mimic Brownian motion, with a temporal resolution of ${\Delta}t=10^{-6}s$. Each spatial jump ${\Delta}R$, of components ${\Delta}X$, ${\Delta}Y$, ${\Delta}Z$, done by the diffusing  particles, depends on the a-priori assigned diffusion coefficient, here $D=408{\mu}m^{2}s^{-1}$, and is \textit{obtained by the composition of three independent displacements that are assumed to be described by random variables with a Gaussian distribution of vanishing mean and standard deviation $\sigma{_x}=\sigma{_y}=\sigma{_z}$}. As $D={\sigma}^2/(6{\Delta}t)$ and ${\Delta}R=\sqrt{{\Delta}X^2+{\Delta}Y^2+{\Delta}Z^2}$, ${\sigma}_x={\sigma}/\sqrt{3}$. The particles are made to move for a time duration $t$, and three vectors of length $l=t/{\Delta}t$ are created with all the random moves realised with Gaussian random variables of vanishing mean and standard deviation $\sigma/\sqrt{3}$. The composition of these vectors thus creates the Brownian path for each particle.
\subsubsection*{Gaussian detection volume}
To mimic FCS experiments we have considered the detection volume as a 3D Gaussian ellipsoid such as:
\begin{equation}\label{Debyelength}
 W(x,y,z) = exp\left({\frac{2(x^2 + y^2)}{r^2_{xy}}} - {\frac{2z^2}{r^2_z}}\right) \;,
\end{equation}
with $x$, $y$ and $z$ the particle position, $r_{xy}$ the minor radius of the confocal volume, and $r_z$ the major radius, with $r_z=k r_{xy}$. To comply with the experimental parameters, we have set $k$=5 and $r_{xy}$=466nm.

The number of photons emitted $n_{ph}$ by a particle at time $t$ and position ($x, y$) in the confocal volume is assumed to be Poisson-distributed. This parameter has been experimentally estimated as $n_{ph}\approx$0.14. However, the parameter $n_{ph}$, determined experimentally, could not be used in the numerical simulations. Indeed, after obtaining the trace (which corresponds to the number of photons detected by the APD as a function of time) its discretization is necessary to mimic the fact that the trace obtained experimentally is discrete. But as the values are relatively close to 0 and 1, the discretization causes important modifications on the correlograms. On the other hand, if the trace is not discretized, the correlograms are not affected by the filters added numerically. Thus the parameter has been empirically set to be  $n_{ph} = 5$ so that the correlograms are not modified by the discretization of the traces, but are still affected by the addition of numeric filters  giving results comparable to the experimental data.
The final result is then an intensity trace coming from the particles passing through the 3D Gaussian ellipsoid and emitting photons with a Poisson distribution.

\subsubsection*{Simulating density filters}
The signal trace is then rounded to obtain a discrete signal with real numbers, then it is filtered.
To mimic the absorption filters, each photon of the signal created previously has a probability $P$ to pass the filter and $1-P$ to be absorbed. We have simulated OD filters with 10$\%$ transmission (OD1, visible in red on Figure \ref{fig:Figure2_FCS-simulations}), 5$\%$ transmission (OD1.3, in green), 1$\%$ transmission (OD2, in blue) and 0.1$\%$ transmission (OD3, in light brown).

\subsubsection*{Data treatment}
The outcomes of numerical simulations of FCS experiments are treated similarly to those obtained in real experiments to work out the autocorrelation function, that is, by adopting
a Fast Fourier Transform algorithm. Thus again:\\
$G(t)=ffti(fft(trace)*Conj(fft(trace)))/(sum(trace)^2-1)$\\
where \textit{trace} now stands for the fluorescence vector generated by the above described numerical simulations of the diffusion of fluorescent molecules, and obtained after the simulated attenuation operated by ideal optical density filters.

\section*{Acknowledgements}
The authors wish to thank S. Mailfert and D. Marguet for helpful discussion and advice. 
The project leading to this publication has received funding from the Excellence Initiative  of Aix-Marseille University - A*Midex, a French ``Investissements d'Avenir'' programme.

\section*{Author contributions statement}
M.L. and M.P. conceived the method. M.L. performed the experiments and data analysis and M.P. supervised the work. The content of this paper stems from part of the PhD thesis work of M.L., who wrote the paper with the contribution of M.P.

\section*{Additional information}
The authors declare to have no competing financial interests.


\bibliography{sample}

\end{document}